\documentclass[journal=jacsat,manuscript=article]{achemso}
\usepackage{amssymb}
\usepackage{soul}
\usepackage{graphicx} 
\usepackage{xcolor}
\usepackage[english]{babel} 
\usepackage{amsmath}
\usepackage{caption}
\usepackage{textcomp}
\usepackage{gensymb}
\usepackage{natbib}
\setkeys{acs}{doi = true}

\title{Chiral TeraHertz surface plasmonics}

\author{Ian Aupiais}
\email{ian.aupiais@unige.ch}
\author{Romain Grasset}
\affiliation[LSI]
{Laboratoire des solides irradi\'{e}s, CEA/DRF/IRAMIS, CNRS, Ecole Polytechnique, Institut Polytechnique de Paris, 91128 Palaiseau, France }
\author{Dmitri Daineka}
\affiliation[LPICM]
{Laboratoire de physique des interfaces et des couches minces, CNRS, Ecole Polytechnique, Institut Polytechnique de Paris, 91128 Palaiseau, France}
\author{Javier Briatico}
\affiliation[THALES]
{Laboratoire Albert Fert, CNRS, Thales, Universit\'{e} Paris-Saclay, 91767 Palaiseau, France}
\author{Luca Perfetti}
\affiliation[LSI]
{Laboratoire des solides irradi\'{e}s, CEA/DRF/IRAMIS, CNRS, Ecole Polytechnique, Institut Polytechnique de Paris, 91128 Palaiseau, France }
\author{Jean-Paul Hugonin}
\author{Jean-Jacques Greffet}
\affiliation[IOGS]
{Universit\'{e} Paris-Saclay, Institut d'Optique Graduate School, CNRS, Laboratoire Charles Fabry, 91127 Palaiseau, France}
\author{Yannis Laplace}
\email{yannis.laplace@polytechnique.edu}
\affiliation[LSI]
{Laboratoire des solides irradi\'{e}s, CEA/DRF/IRAMIS, CNRS, Ecole Polytechnique, Institut Polytechnique de Paris, 91128 Palaiseau, France }

\begin{document}

\begin{abstract}
 Chiral engineering of TeraHertz (THz) light fields and the use of the handedness of light in THz light-matter interactions promise many novel opportunities for advanced sensing and control of matter in this frequency range. Unlike previously explored methods, this is achieved here by leveraging the chiral properties of highly confined THz surface plasmon modes. More specifically, we design ultrasmall surface plasmonic-based THz cavities and THz metasurfaces that display significant and adjustable chiral behavior under modest magnetic fields ($B\leq500mT$). For such a prototypical example of non-hermitian and dispersive photonic system, we demonstrate the capacity to magnetic field-tune both the poles and zeros of cavity resonances, the two fundamental parameters governing their resonance properties.
 Alongside the observed handedness-dependent cavity frequencies, this highlights the remarkable ability to engineer chiral and tunable radiative couplings for THz resonators and metasurfaces. The extensive tunability offered by the surface plasmonic approach paves the way for the development of agile and multifunctional THz metasurfaces as well as the realization of ultrastrong chiral light-matter interactions at low energy in matter with potential far-reaching applications for the design of material properties.
\end{abstract}

\section*{Introduction}
Despite its long history, the role of chirality in light-matter interactions remains intriguing, drawing intense focus both from a fundamental perspective \cite{Tang_2010,Mun_2020} and for its technological applications in sensing and controlling matter across diverse length scales, energy scales and scientific disciplines \cite{Tang_2011, Bord_cs_2012, Lodahl_2017,Brandt_2017,Lininger_2022,Genet_2022}. In this context, the design of chiral optical fields and optically-active devices at THz frequencies holds particular significance for sensing applications across a large class of systems. This stems from the wide range of excitations exhibiting chirality within this frequency band, spanning from chemically and biologically relevant excitations in molecules \cite{Choi_2022,Choi2022} to quantum collective modes such as magnons \cite{Bord_cs_2012, Kampfrath_2010,Cui_2023} and phonons \cite{Cui_2023,Ishito_2022,Baydin_2022} in condensed matter systems. Alongside sensing capabilities, the ultrastrong interaction between matter and chiral optical fields at low energies has recently emerged as a new paradigm for manipulating material properties through Floquet engineering \cite{Wang_2013,McIver_2019,Topp_2019} and has been proposed to realize novel phases of matter through hybridization with the vacuum field of chiral cavities \cite{Wang_2019,H_bener_2020, Tokatly_2021,Sedov_2022,Owens_2022}. Notably, the potential to induce magnetism and non-trivial topologies in otherwise non-magnetic and topologically trivial materials \cite{Wang_2019,Sedov_2022, Masuki_2023}, or photon condensation in the ground state \cite{Mercurio_2024}, represent some of the most far-reaching prospects enabled by chiral vacuum fields. For these reasons, the ability to engineer ultrastrong and chiral light-matter interactions at THz frequencies, achievable through the design of cavities with ultrasmall mode volumes, appears as a new frontier in the THz range \cite{Andberger_2023} that would open up many avenues for chiral sensing and matter manipulation with applications in chemistry, biology and physics.\\
In the past years, large efforts have been devoted to the design of chiral and optically-active devices for polarization control of THz light \cite{Zhang_2012, Wu_2014, Kan_2015,Kim_2017,Choi_2019,Cong_2019, Mu_2019, Tan_2021, Ju_2023}. 
These have been obtained following mostly two distinct routes. The first one is a synthetic or metamaterial-like approach, in which the geometric chirality of metallic resonators is tailored and transferred onto the resonating electromagnetic (EM) fields \cite{Zhang_2012, Wu_2014, Kan_2015,Kim_2017,Choi_2019,Cong_2019}. The second approach relies on the natural magneto-optical activity of the bulk of semiconductors \cite{Mu_2019, Tan_2021, Ju_2023} and is enabled by the propagation of THz magneto-plasma waves \cite{Palik_1970, Wang_2009}. The latter originate from the microscopic Lorentz force acting on carriers with low effective mass $m^{*}$, giving rise to a cyclotron frequency $\nu_c=eB/2\pi m^{*}$ that falls within the THz frequency range for moderate magnetic fields below 1T \cite{Shimano_2002}. Yet, achieving high EM-field confinement sufficient enough to reach the ultrastrong coupling regime of light-matter interaction represents to date a challenge for both of these approaches. On the one hand, shrinking the size of metallic resonators while preserving operating frequencies in the THz range is a difficult task that often requires the design of complex and technologically challenging resonator architectures \cite{Paulillo_2014, Keller_2017, Mottaghizadeh_2017}. On the other hand, utilization of magneto-plasma waves necessitates light propagation through the bulk of the semiconductor and hence only offer sizeable chiral effects at the expense of large interaction volumes \cite{Ju_2023,Wang_2009}. \\
Here, we overcome these challenges by exploiting THz surface plasmon-based cavities in semiconductors \cite{Aupiais_2023}, an approach which allows to combine the exceptional confinement ability of surface plasmons with the tunable magneto-plasmonic properties of semiconductors at THz frequencies in order to functionalize EM-fields and cavity properties at deep-subwavelength scales. Specifically, we show that surface plasmons inherit the chiral activity of the bulk and use this as a means to create chiral THz cavities with ultrasmall mode volumes. When assembled together, arrays of such cavities lead to a novel type of chiral THz metasurfaces, fully tunable by geometry, magnetic-field and even temperature. This work opens the door to exciting perspectives for the exploration of ultrastrong chiral light-matter interactions at low energies in matter.

\section*{Working principle of the chiral THz surface plasmonic cavities}
\begin{figure*}[!ht]
	\centering 
	\includegraphics[width=\textwidth]{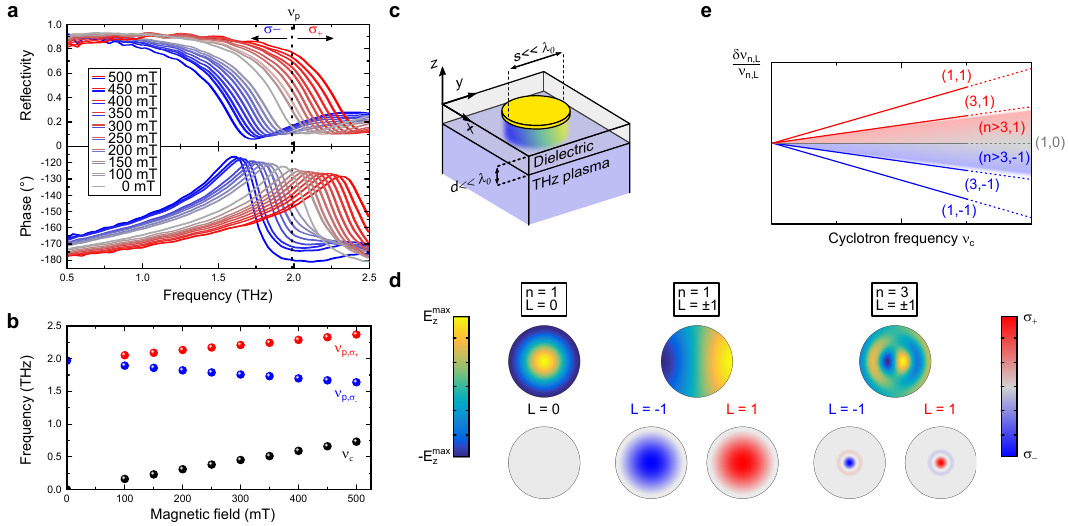}
	\caption{\textbf{a)} ${\sigma }_{\pm }\ $polarization-resolved power spectra (top panel) and phase (bottom panel) of the reflected THz field from bulk InSb at various magnetic fields between $0mT$ and $500mT$. Measurements are performed at normal incidence and the magnetic field is applied perpendicular to the surface of the plasmonic medium. The dashed line indicates the location of plasma frequency $\nu_p$ at $B=0 mT$. \textbf{b)} Cyclotron frequency (black dots) as a function of magnetic field, determined from fitting the bulk reflectivity spectra with the magneto-plasmonic model of permittivities (see also supporting material). Red and blue dots indicate handedness-dependent plasma frequencies ${\nu }_{p,{\sigma }_{\pm }}$ associated with the cyclotron resonance active and inactive  bulk magneto-plasma waves (\cite{Palik_1970}) \textbf{c)} Sketch of a single subwavelength chiral THz plasmonic cavity with overall dimensions $s,d\ll \lambda_0 $. \textbf{d)} Characteristics of the plasmonic modes for some selected cavity resonances of the circular patch cavity without a magnetic field: $(n,L)=\{\left(1,0\right),(1,+1),\ (1,-1),\ (3,+1),\ (3,-1)\}$. Shown are color plots of: 1) the normal component of the EM-field ($Re(\tilde{E}_z)$) (\textit{top}) and 2) relative density of the ${\sigma }_{\pm }$ polarized photons (or 'spin' density, bottom) of the corresponding surface plasmon \textbf{e)} Resonance splitting expected for the different orbital modes of the subwavelength cavity as a function of the cyclotron frequency.}
	\label{fig:1}
\end{figure*}

In the Faraday configuration, where a homogeneous magnetic field $\boldsymbol{B}=B\boldsymbol{z}$ is perpendicular to the plasmonic medium's interface and a plane wave is incident normally, the magnetic field acts exclusively onto the spin-sector of the EM-field (i.e. its polarization state) and normal modes correspond to left- and right-circularly polarized (LCP/RCP) photons
${\widetilde{\boldsymbol{E}}}_{{\sigma }_{\pm }}\boldsymbol{=}e^{ik_zz}e^{-i\omega t}\left.\left|{\sigma }_{\pm }\right.\right\rangle $  \cite{Palik_1970}. Interaction between circularly-polarized light and the plasma involves two distinct magneto-plasmonic permittivities which read in the circular basis ${\epsilon }_{\sigma_{\pm} }\left(\nu,B \right)={\epsilon }_{xx}\left(\nu ,B\right)\pm i{\epsilon }_{xy}\left(\nu,B \right)={\epsilon }_{\infty }\left(1-\frac{{{\nu }_p}^2}{\nu }*\frac{1}{\nu \pm {\nu }_c+i\mathrm{\Gamma }}\right)$, where $\nu$, $\nu_p$, $\nu_c$, $\Gamma$ and $\epsilon_\infty$ are the frequency, plasma frequency, cyclotron frequency, carrier scattering rate and background permittivity of the material, respectively (see Supporting Information). Accordingly, this results in two distinct reflection coefficients denoted below as $\tilde{r}_{{\sigma }_\pm}(\nu,B)$ for $\sigma_\pm$ polarized light. \\
\noindent Normal incidence experimental power reflectivity spectra $R_{{\sigma }_\pm}(\nu)=\arrowvert \tilde{r}_{{\sigma }_\pm}(\nu,B)\arrowvert^2$ as well as the phase $\phi_{{\sigma }_\pm}(\nu,B)=Arg(\tilde{r}_{{\sigma }_\pm}(\nu,B))$ accumulated by the electric field of the light upon reflection onto the THz plasma of bulk semiconductor InSb are shown in Fig. 1a for $\sigma_\pm$ polarizations. As the cyclotron frequency ${\nu }_c$ varies, the THz plasma edge initially located at $\nu \approx {\nu }_p$ splits into two different plasma edges depending on the polarization handedness. The corresponding plasma frequencies, defined by the zero-crossing of the two permittivities ${\epsilon }_{\sigma_{\pm} }\left(\nu,B \right)=0$, amount to ${\nu }_{p,{\sigma }_{\pm }}\approx \frac{1}{2}\left(\sqrt{{4\nu }^2_p+{\nu }^2_c}\pm {\nu }_c\right)$ in the limit of small carrier scattering rate and evolve oppositely as a function of $\nu_c$ (see Fig. 1b), highlighting the bulk chirality of the plasmonic medium\cite{Palik_1970,Wang_2009}.\\ 
\noindent In contrast to plane waves, where orbital and spin components of the EM-field are independent, the normal modes of the proposed ultrasmall chiral THz cavity illustrated in Fig. 1c are sustained by surface plasmons and their orbital and spin degrees of freedom are intertwined within the highly subwavelength volume of the resonator, leading to a more complex chiral behavior. This can be seen in Fig. 1d where we present the characteristics of the normal modes of circular patch cavities, which, owing to their cylindrical symmetry, represent a natural choice of plasmonic resonators in the present context. In the absence of magnetic field, their normal modes have been studied previously \cite{Minkowski_2014} and are characterized by an orbital angular moment $L\ (L=\ \dots ,-2,-1,0,1,2,\dots )\ $ and a radial number $n\ (n=1,2,\ \dots )$. For lossless cavities, modes of opposite orbital momentum are time-reversal symmetric of one another (${\widetilde{\boldsymbol{E}}}_{n,-L}\boldsymbol{=}{\widetilde{\boldsymbol{E}}}^{\boldsymbol{*}}_{n,L}$) and hence have degenerate frequencies ${\nu }_{n,L}={\nu }_{n,-L}$. In Fig 1d, we display the EM-fields sustained by the surface plasmon inside the resonator for a set of representative $(n,L)$ modes: 1) the normal component of the electric field  ($Re(\tilde{E}_z)$)  and 2) the relative density $\mathrm{\Delta }{\sigma }_{n,L}\left(r\right)={\sigma }_{+,n,L}\left(r\right)-{\sigma }_{-,n,L}\left(r\right)$ of the ${\sigma }_{\pm }$ polarized photons (or spin density of the mode), where ${\sigma }_{\pm,n,L}\left(r\right)\propto \mid \tilde{E}_x \pm i\tilde{E}_y \mid ^2$ depends only on the radial coordinate $r$. 

\noindent Notably, normal modes are seen to consist in a spatially dependent superposition of ${\sigma }_{\pm }$ polarized photons which entails the mixing of orbital and spin degrees of freedom of the resonator. It is possible to estimate the shifts $\delta {\nu }_{n,L}$ of the resonance frequencies ${\nu }_{n,L}$ by applying a modal perturbation theory, which, to first order in ${\nu }_c$, gives (see Supporting Information):

\noindent 
\begin{equation}
\label{eq1}
\frac{\delta {\nu }_{n,L}}{{\nu }_{n,L}}\propto {\nu }_c\left(\int^{s/2}_0{r{\mathrm{\ }\mathrm{\Delta }\sigma }_{n,L}\left(r\right)dr}\right)
\end{equation}

\noindent indicating that the frequency shift is proportional to the spatially integrated spin density ${\mathrm{\ }\mathrm{\Delta }\sigma }_{n,L}\left(r\right)$ of the ${\sigma }_{\pm }$ polarized photons ($s$ being the diameter of the patch). As depicted in Fig. 1e, the $L=0$ mode doesn't experience any shift as it is linearly polarized, while modes with finite orbital angular momentum $(L\neq 0)$ split symmetrically according to $\delta {\nu }_{n,L}=-\delta {\nu }_{n,-L}$ as they have opposite spin densities of equal magnitude. Because they carry the largest spin polarization among all, the $(n=1, L=\pm 1)$ modes are expected to provide the most pronounced mode splitting and hence the largest chiral effect. 

\section*{Sample architecture and plasmonic resonances without magnetic field}

\noindent Rather than an isolated cavity, our samples consist in two-dimensional and subwavelength periodic arrays (period $s+a$) of circular patch cavities of diameter $s\ $(Fig. 2a). Over a bulk wafer of semiconducting InSb, we deposit an insulating layer of highly subwavelength thickness $d$ (${Si}_{3}{N}_{4}$ with $d\ll \lambda_0 $, $\lambda_0 $ being the wavelength in vacuum) followed by circular patches of metallic gold on top. In the absence of magnetic field and due to the four-fold symmetry of the structure, reflectivity spectra are independent of the incoming THz light polarization.  In Fig. 2b, we present the experimental and simulated power reflectivity spectra for three samples having circular patches of varying diameters. The major absorption observed at low frequency corresponds to the resonance frequency of the two degenerate $(n,L)=(1,\pm 1)$ cavity modes, as confirmed by inspection of the mode profile obtained from numerical simulations (Fig.  2c). A second pronounced absorption at high frequency lies slightly below the THz plasma edge. As seen in Fig. 2c, it corresponds to the resonance of a pair of degenerate plasmonic modes that are fully expelled from the volume encompassed by the cavities and that are located between them. These modes are the counterpart in these structures of the inter-cavity mode observed in striped lattices of plasmonic cavities \cite{Aupiais_2023}. For completeness, we also report the presence of: 1) a small absorption for the $s=43\mu m$ sample (triangle in Fig. 2b) which corresponds to the excitation of the higher order $(n,L)=(3,\pm 1)$ cavity modes and whose mode profile is shown in Fig.2c and 2) a resonance corresponding to the excitation of a continuum of propagating bulk plasma modes within the transparency region of the plasmonic medium arising the first-order diffraction of the array ($s=43\mu m$ sample, star in Fig. 2b). In the remaining of this paper, we will not further discuss the latter resonances and will focus our analysis on the main cavity resonance. 
\begin{figure*}[!ht]
	\centering 
	\includegraphics[width=\textwidth]{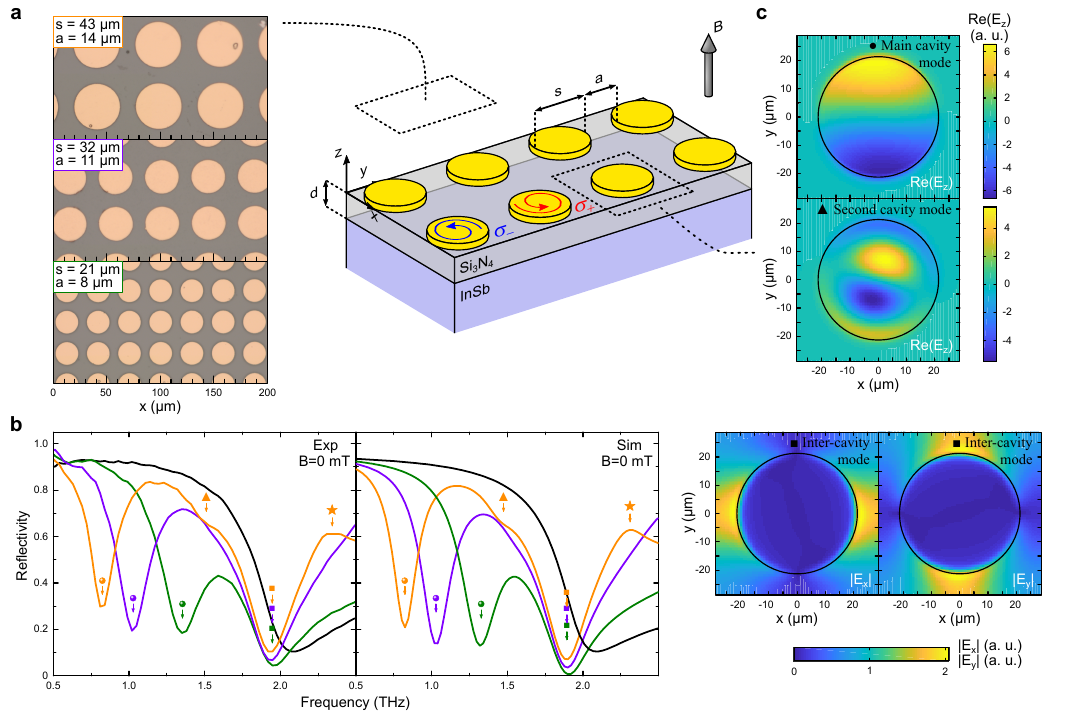}
	\caption{\textbf{a)} Sample architecture and geometrical parameters of the two-dimensional array of circular patch cavities (right). Top-view optical pictures of the actual samples (left). \textbf{b)} Experimental (left) and simulated (right) normal incidence THz power reflectivity spectra of the 3 samples investigated at zero magnetic field ($B=0mT$): $s=43\mu m$ (orange solid curve), $s=32\mu m$ (purple solid curve), $s=21\mu m$ (green solid curve). The normal incidence THz power reflectivity spectrum of bulk InSb is reported for comparison (black solid curve). The different plasmonic resonances are reported: main cavity resonance (dots), second cavity resonance (triangle), intercavity resonance (squares), first-order diffraction of the array (star) \textbf{c)} Simulated EM mode profiles of the $s=43\mu m$ sample performed at frequencies corresponding to the resonances indicated in panel b. All mode profiles are shown for a $z-$cut taken in the middle of the dielectric insulator at $z=d/2$. Top: Mode profile of the main cavity resonance (plot of $Re(\tilde{E}_z)$). Center: Mode profile of the second cavity resonance (plot of $Re(\tilde{E}_z)$). Bottom: Mode profile of the inter-cavity resonance (plots of $|\tilde{E}_x|$ and $|\tilde{E}_y|$)}
	\label{fig:2}
\end{figure*}

\section*{Surface plasmonic chirality and its magnetic field dependence}

\noindent Upon application of the magnetic field, we observe the lifting of the degeneracy of the $\left(n,L\right)=(1,+1)$ and $(1,-1)$ cavity resonance. This is featured in the polarization resolved THz spectra presented in Fig. 3a-b-c as two frequency-displaced resonant absorptions for opposite ${\sigma }_{\pm }$ polarizations, hence demonstrating the chiral effect that is looked for.
\begin{figure*}[h!]
	\centering 
	\includegraphics[width=\textwidth]{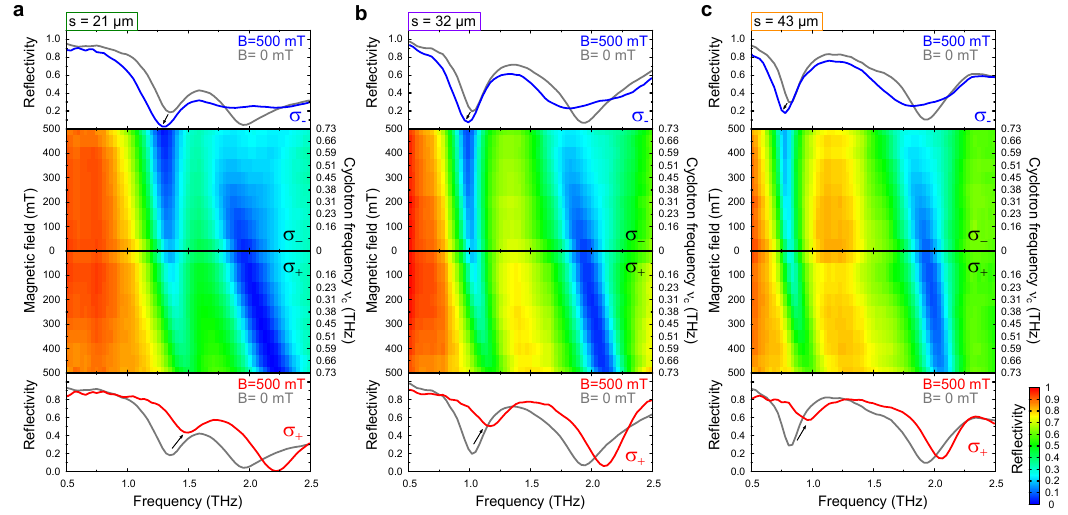}
	\caption{Experimental measurements of the ${\sigma }_{\pm }$ polarization-resolved THz power reflectivity spectra as a function of magnetic field between $0mT$ and $500mT$ for the three samples investigated: \textbf{a)} $s=21 \mu m$ \textbf{b)} $s=32 \mu m$ and \textbf{c)} $s=43 \mu m$. The central panels display the THz reflectivity spectra in color scale. For each magnetic field, the corresponding cyclotron frequency is reported on the right axis. The top and bottom panels show the THz spectra at $B=500mT$ obtained for the ${\sigma }_+$ (bottom panel, red solid curves) and ${\sigma }_-$ (top panel, blue solid curves) polarizations, together with the THz spectrum obtained at $B=0mT$ for comparison (grey solid curves).}
	\label{fig:3}
\end{figure*}
\noindent Besides the polarization handedeness-dependent frequency-shifts, we notice that the contrast of the cavity resonance varies strongly and evolves oppositely for opposite ${\sigma }_{\pm }$ polarizations. The large variation of its contrast can be inferred from the value of the reflectivity minimum at resonance $R_{min}$ which changes as much as $\sim $50\% within the range of parameters investigated (see top and bottom panels in Fig. 3a-b-c). \\
\noindent This observation suggests that the magnetic field is acting to tune another crucial cavity parameter: the overall coupling efficiency of external radiation to the cavity. It can be understood qualitatively from temporal coupled mode theory \cite{Fan_2003} which predicts that $R_{min}={({\mathrm{\Gamma }}_r-{\mathrm{\Gamma }}_{\ohm})}^2/{({\mathrm{\Gamma }}_r+{\mathrm{\Gamma }}_\ohm)}^2$ for single-port cavities \cite{Manceau_2013,Qu_2015}, where ${\mathrm{\Gamma }}_r$ and ${\mathrm{\Gamma }}_\ohm$ are the radiative and non-radiative decay rates of the cavity describing respectively the coupling of the resonator to external radiation and the decay of the resonance due to ohmic losses. In other words, $R_{min}$ measures how close the cavity is from critical coupling at which ${\mathrm{\Gamma }}_\ohm={\mathrm{\Gamma }}_r$ ($R_{min}=0$) and all incident radiation is coupled and dissipated into the cavity. In order to further assess quantitatively and reliably all cavity parameters for this highly non-Hermitian and dispersive photonic system \cite{Alpeggiani_2017,Zhang2020,Benzaouia_2021}, we fitted the two quadratures of the reflection coefficient $\tilde{r}_{{\sigma }_\pm}(\nu,B)$ based on an analytical continuation of the frequency in the complex plane \cite{Popov_1986,Nevi_re_1995,Binkowski_2024} ($\nu \rightarrow\tilde{\nu}$ and $\tilde{r}_{{\sigma }_\pm}(\nu,B)\rightarrow \tilde{r}_{{\sigma }_\pm}(\tilde{\nu},B)$, see Supporting Information). This leads to a pole-zero representation of the cavity resonance shown in Fig. \ref{fig:4}a. We define the complex zero $\tilde{\nu}_{z0}$ by the condition  $\tilde{r}_{{\sigma }_\pm}(\tilde{\nu}_{z0})=0$.  Following this zero in the complex plane enables to characterize the undercoupled ($Im(\tilde{\nu}_{z0})<0)$), critically coupled ($Im(\tilde{\nu}_{z0})=0)$ and overcoupled $(Im(\tilde{\nu}_{z0})>0)$ regimes. Note that the critically coupled regime is closely linked to the coherent perfect absorption \cite{Chong_2010,Wan_2011,Sweeney_2020}. We observe that as the magnetic field is raised, the zero either moves towards ($\sigma_{-}$) the real axis (critical coupling) or departs from it ($\sigma_{+}$), signaling not only the tunability of the coupling but also its chiral behavior. The major results of the analysis of the cavity parameters are further presented in Fig. \ref{fig:4}b.\\
\begin{figure*}[!ht]
	\centering 
	\includegraphics[width=\textwidth]{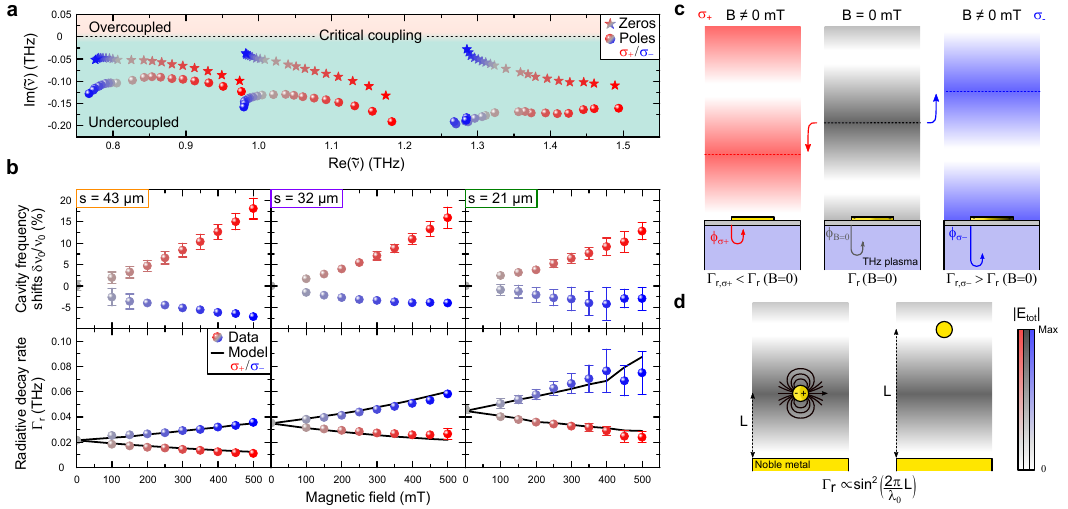}
	\caption{\textbf{a)} Location of the poles ($\tilde{\nu}_0$) and zeros ($\tilde{\nu}_{z0}$) of the cavity resonance in the complex plane and their trajectory  under application of the magnetic field. \textbf{b)} Chirality of the cavity parameters as a function of magnetic field for $\sigma_{+}$ (red dots) and $\sigma_{-}$ (blue dots) polarizations. Top panels: relative frequency-shifts $\delta{\nu }_{0,\sigma_{\pm}}/{\nu}_{0} $. Bottom panels: radiative decay rates ${\mathrm{\Gamma }}_{r,\sigma_{\pm}}$. The experimental data (dots) is compared to the model describing the variation of the radiative decay rates in a magnetic field (black solid lines): ${\mathrm{\Gamma }}_{r,\sigma_{\pm}}(B)=\alpha \left| 1+\tilde{r}_{\sigma_{\pm},0}(\nu,B) \right| ^2 $(see main text). From left to right: $s=43, 32$ and $21 \mu m$ cavity samples. \textbf{c} Sketch of the origin of the chirality and of the tuning of the radiative decay rates with the magnetic field. In the absence of the gold patch, incoming and reflected light from the plasmonic medium interfere to form standing waves with total amplitude  $\left| E_{tot,\sigma_{\pm}}\right|$ (the background field) whose location of nodes and anti-nodes depend on the phase $\phi_{\sigma_{\pm}}(\nu,B)$ imposed by the medium upon reflection, as shown by the dashed lines which indicate the location of the anti-nodes of the background field. When the gold patch is then introduced, it acts both as a support to the cavity resonance and as a coupler to it. Its location with respect to the background field determines how much light is coupled in and out of the cavity. The chirality and tunability of the phase $\phi_{\sigma_{\pm}}(\nu,B)$ of the bulk plasmonic medium, as shown in Fig. 1a, is hence rooting that of the radiative decay rates. \textbf{d} Analogy of the tuning of the radiative decay rates with a resonant nanoparticle located above a metallic mirror. In the absence of the nanoparticle, incoming and reflected light from the metallic mirror form standing waves. When the nanoparticle is then immersed into this background field, its location determines the amount of light that is scattered and used to drive its resonance, i.e. the radiative decay rates. If it is located at the anti-node of the background field, the absence of light inhibits the resonance, while if it is located at a node, the resonance is maximally driven.  The height of the nanoparticle controls in this case the radiative decay rates and tunability is achieved geometrically.}
	\label{fig:4}
\end{figure*}
\noindent The relative frequency-shifts of the cavity resonance are defined as $\frac{\delta {\nu }_{0}}{{\nu }_{0}} = \frac{{\nu }_{0,\sigma_{\pm}}(B)-{\nu }_{0}(B=0)}{{\nu }_{0}(B=0)}$ where ${\nu }_{0,\sigma_{\pm}}(B)=Re(\tilde{\nu }_{0,\sigma_{\pm}}(B))$ corresponds to the real part of the cavity pole $\tilde{\nu }_{0,\sigma_{\pm}}(B)$, i.e. the cavity resonance frequency. As anticipated, we observe that they demonstrate opposite variations for $\sigma_{\pm}$ polarizations as a function of magnetic field $B$. While linear in $B$ at small fields, they acquire a slight quadratic dependence at larger fields which is reminiscent of the dependences of bulk magneto-plasma modes \cite{Wang_2009}. At $B=500 mT$, the relative shifts amount to 14$\%$, 20$\%$ and 24$\%$ for the $21, 32$ and $43\mu m$ cavities respectively in between the two $\sigma_{\pm}$ polarizations. The two decay rates, determined from the knowledge of the pole and the zero as $\Gamma_{\ohm}=-Im(\tilde{\nu}_0+\tilde{\nu}_{z0})/2$ and $\Gamma_{r}=-Im(\tilde{\nu}_0-\tilde{\nu}_{z0})/2$ (see Supporting Information), also evolve as a function of magnetic-field and exhibit chiral behaviour: ${\mathrm{\Gamma }}_{r/ \ohm,\sigma_{+}}(B)\neq{\mathrm{\Gamma }}_{r/ \ohm,\sigma_{-}}(B)$. While this is somewhat expected for the ohmic decay rate as the cyclotron resonance modulates material dissipation in the vicinity of $\nu=\pm \nu_c$, the most striking feature is the chiral and large variations observed for the radiative decay rate by as much as 300$\%$ in the range of parameters investigated. Such phenomenon is largely unexpected as this quantity is generally believed to depend solely on geometrical parameters of the arrays \cite{Manceau_2013, Qu_2015, Feuillet_Palma_2013}. We attribute it to the magnetic field-tuning of the phase $\phi_{\sigma_{\pm}}(\nu,B)$ accumulated by the THz fields upon reflection onto the underlying plasmonic medium. More specifically, denoting $\tilde{r}_{\sigma_{\pm},0}(\nu,B)$ the background reflectivity of the structure \textit{in the absence of metallic patches} (i.e. involving only the $vacuum/Si_3 N_4/InSb$ planar multilayer), we show that the variation of the radiative decay rate can be well reproduced and explained by a functionnal dependence of the form ${\mathrm{\Gamma }}_{r,\sigma_{\pm}}(B)=\alpha \left| 1+\tilde{r}_{\sigma_{\pm},0}(\nu,B) \right| ^2 $ where $\tilde{r}_{\sigma_{\pm},0}(\nu,B)$ is evaluated at the $vacuum/Si_3 N_4$ interface, at the cavity resonance frequency $\nu=\nu_{0,\sigma_{\pm}}(B)$, and $\alpha$ is a geometry-dependent but magnetic field-independent scaling factor (black solid lines in Fig. 4b). Such a relation between background reflectivity and radiative decay rates has been derived analytically for metal/insulator/metal cavities in a context where the phase did not play any significant role \cite{Bowen_2016,Bowen_2017}. In the present case however, the phase $\phi_{\sigma_{\pm}}(\nu,B)$  of the reflected field is the central parameter controlling such behavior, as within a good approximation $\left|\tilde{r}_{\sigma_{\pm},0}(\nu,B) \right| \approx 1$, so that ${\mathrm{\Gamma }}_{r,\sigma_{\pm}}(B) \propto \cos(\phi_{\sigma_{\pm}}(\nu,B)/2)^2 $. The basic idea behind the tunability and chirality of the radiative decay rates that we observe is sketched in Fig. 4c. In the absence of metallic patches, interference of the incident $E_{i,\sigma_{\pm}}$ and reflected fields $E_{r,\sigma_{\pm}}=\tilde{r}_{\sigma_{\pm},0}(\nu,B)E_{i,\sigma_{\pm}}$ result in a total background field of amplitude $\left| E_{tot,\sigma_{\pm}}\right|=\left| 1+\tilde{r}_{\sigma_{\pm},0}(\nu,B)\right| \left| E_{i,\sigma_{\pm}}\right| $ that takes the form of a standing wave whose locations of nodes and anti-nodes are crucially depending the phase $\phi_{\sigma_{\pm}}(\nu,B)$ accumulated upon reflection from the plasmonic medium. Immersed into this background field, the metallic patches in- and out- couple it to the cavity with coupling rates $\sqrt{2\mathrm{\Gamma }_{r,\sigma_{\pm}}(B)}$ \cite{Fan_2003} being determined by the field amplitude at their location and hence scaling as $ \left| 1+\tilde{r}_{\sigma_{\pm},0}(\nu,B) \right|$. This leads to an overall radiation decay rate given by: ${\mathrm{\Gamma }}_{r,\sigma_{\pm}}(B) \propto \left| 1+\tilde{r}_{\sigma_{\pm},0}(\nu,B) \right| ^2 $. This phenomenon is analogous to a resonant nanoparticle located at distance $L$ above a metallic mirror and whose coupling to external radiation ${\mathrm{\Gamma }}_r\propto\sin(2\pi L/\lambda_0)^2$ is adjustable by varying the location of the nanoparticle with respect to the standing wave (see Fig. 4d). Here, despite the fixed geometry, the sizeable variations of the phase provided by the underlying plasmonic medium close to the plasma frequency as well as their tunability with magnetic field play the same role as the height for the nanoparticle. \\
\noindent As such, for the proposed plasmonic cavities, we demonstrate that chirality proceeds not only by frequency splitting of degenerate modes but also by a chiral and tunable coupling of the resonators to external radiation. This constitutes an important result of this study.
\noindent 
\section*{Chiral THz surface plasmonic metasurfaces}

\noindent We discuss here an application where the proposed structures can be exploited as resonant and tunable reflective THz chiral metasurfaces for polarization control of the THz light. In order to demonstrate the advantages of structuring a metasurface over that of using a bulk semiconductor for this purpose, we compare in Fig. 5 the optical activity of the unstructured bulk material and that of a structured metasurface. Optical activity is commonly characterized in terms of: 1) circular dichroism (CD) via the ellipticity $\eta=\frac{|r_{\sigma_{+}}(\nu)|-|r_{\sigma_{-}}(\nu)|}{|r_{\sigma_{+}}(\nu)|+|r_{\sigma_{-}}(\nu)|}$ and 2) optical rotary dispersion (OR) via the rotation angle $\theta=\frac{\phi_{\sigma_{+}}(\nu)-\phi_{\sigma_{-}}(\nu)}{2}$\cite{Zhou_2012}.  For linearly polarized incident light, CD and OR result in the present context in reflected light that is elliptically polarized (CD phenomenon) and rotated by an angle $\theta$ compared to the incident polarization (OR phenomenon), see Fig. 5a.
In Fig. 5b, we compare the ellipticity and optical rotation angle of the bulk and structured metasurface at $B=500mT$.\\
\begin{figure*}[ht]
	\centering 
	\includegraphics[width=\textwidth]{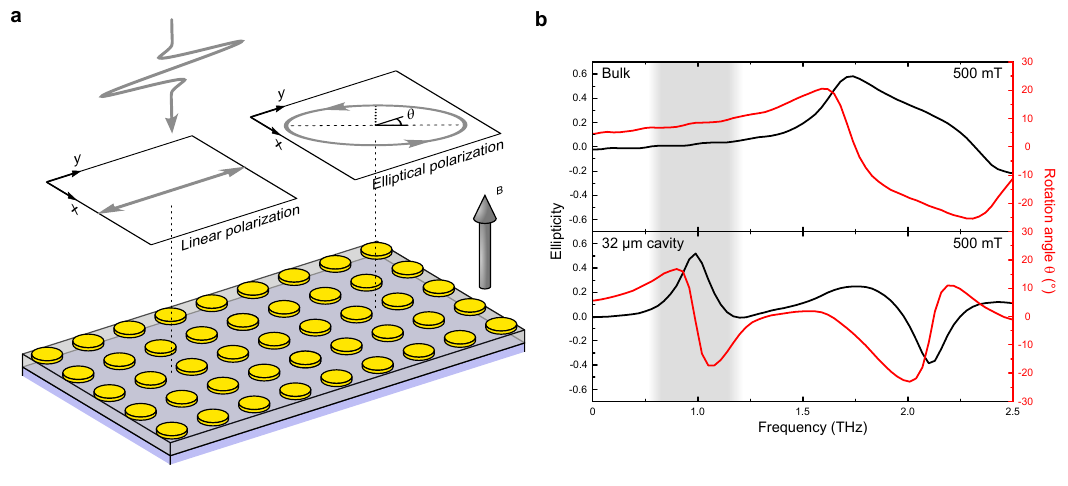}
	\caption{Comparison between the optical activity arising from the bulk semiconductor and a THz metasurface. \textbf{a} Sketch of the change in the polarization of the light as due to optical activity from either the bulk semiconductor or a structured metasurface (here shown is the metasurface). Optical activity manifests in circular dichroism (CD) and optical rotary dispersion (OD). For linearly polarized incident light, CD results in elliptical polarization while OR results in optical rotation of the polarization by an angle $\theta$ of the reflected light. \textbf{b} Comparision of CD and OR between the bulk semiconductor (top panel) and $s=32 \mu m$ metasurface (bottom panel) at $B=500 mT$. }
	\label{fig:5}
\end{figure*}
\noindent We notice profound differences between the two responses. Firstly, we observe that CD and OR of the bulk material display rather complex shapes in an otherwise large frequency window centered around the plasma edge. This results in overall poor spectral control of the optical activity of the bulk material. In contrast, the ellipticity of the metasurface presents two sizeable peaks of opposite polarity at around 1THz and 2THz, which correspond to the frequencies of the cavity and intercavity resonances respectively for this $s=32 \mu m$ cavity metasurface. Ellipticity around the cavity resonance demonstrates in particular a well-defined unipolar behaviour.\\
Strikingly, at the cavity resonance, the ellipticity  reaches a sizeable value $\eta \approx 0.5$ that is at a frequency where the bulk ellipticity is almost null within noise level: this corresponds to a giant enhancement of the ellipticity provided by the metasurface. The ability to induce such a large dichroism in a frequency band well below the plasma frequency where the bulk material only provides a minute effect demonstrates the superiority of metasurface over the bulk material for the chiral engineering of THz light fields. The origin of such a metasurface-enabled ellipticity can be traced back to the difference in the contrast between  $\sigma_{+}$ and $\sigma_{-}$ resonances. In this respect, chirality of the radiative decay rates of the cavities discussed in the previous section is crucial in providing this effect. \\
\noindent In the Supporting Information, we further show the wide tunability of the optical activity of the metasurfaces both spectrally via the geometrical tuning of the cavity dimensions and in its magnitude via the tuning of the magnetic field strenght. Such metasurfaces achieve ellipticities as high as $\eta \approx 0.7$ at a moderate field $B=500mT$ relevant for applications, and we expect that full ellipticity ($\eta=1$) is reachable around $600-700mT$. 

\section*{Conclusion}
Our approach offers a major and unique asset with respect to other alternatives for the chiral engineering of THz fields based on either synthetic \cite{Zhang_2012,Kan_2015,Kim_2017,Cong_2019} or natural \cite{Wang_2009,Mu_2019,Ju_2023} optical activity. In comparison, the large magnitude of optical activity obtained here is achieved over much smaller, i.e. deep-subwavelength, scales thanks to the remarkable confinement properties of surface plasmons. All the reported chiral effects originate from beneath the surface of the plasmonic medium within the penetration depth of the surface plasmons that is $\sim 0.01 \lambda_0$ (where $\lambda_0$ is the free space wavelength) and within cavities with mode volumes of the order of $10^{-6} \lambda_0^3$. This represents at least two orders of magnitude enhancement over previous approaches for typical figure of merits that compare the magnitude of CD/OR and the lenght or volume over which it is realized. Straightforward improvements can be readily achieved by reducing cavity mode volumes to $10^{-8} \lambda_0^3$ \cite{Aupiais_2023} or by increasing the ratio $\nu_c/\nu_p$ through the magnetic field or the temperature \cite{Aupiais_2023}, as this ratio governs the overall magnitude of the chiral effects. In summary, along with the numerous opportunities it allows for harnessing ultrastrong chiral light-matter interactions for the sensing and control of matter at THz frequencies, this system stands out as a singular platform within the electromagnetic spectrum for unravelling novel phenomena in chiral plasmonics thanks to the extended parameter space in $\nu_c/\nu_p$ which remains beyond reach at mid-infrared and optical frequencies.

\section*{Methods}

\subsection*{Sample manufacturing}

Samples were manufactured starting from a $500 \ \mu m$ thick, $<100>$ oriented, bulk wafer of InSb commercially available (MTI Corporation). The InSb wafer was nominally N-type undoped. Each sample consisted of a $\sim 5 \ mm*5 \ mm*0.5 \ mm$ InSb substrate dice-cut from the wafer. An insulating layer of Si$_3$N$_4$ was deposited via plasma enhanced chemical vapor deposition and the top metallic patches were realized via photolithography followed by metal deposition of Ti(15 nm)/Au(200 nm). The geometric and optical parameters of the samples are provided in the Supporting Information.

\subsection*{THz spectroscopy}

Spectroscopy of the samples was performed with a time-domain THz spectrometer driven by an ultrafast Ti:Sapphire oscillator. THz generation and detection were achieved via a photoconductive emitter (Tera-SED) and a 1 mm thick $<110>$ ZnTe crystal, respectively, allowing for spectroscopic coverage from 0.2 THz to 2.5 THz. Delayed THz pulses originating from the optical components of the setup were eliminated by windowing the main THz pulse reflected off the samples. The time domain signals were Fourier transformed and analyzed in both their amplitude and phase. All reflectivity measurements were conducted at normal incidence and room temperature.  For absolute measurements of the reflectivity, we used the THz pulse reflected off a gold-coated sample as a reference. In all measurements, the THz spot size ($\sim 2 \ mm$ in diameter) is well below the size of the manufactured samples, so that the samples can be considered infinite in the plane containing the interface ($x$-$y$ plane in Fig. \ref{fig:2}.a).
Polarization-resolved THz measurements were obtained by initially decomposing the linearly-polarized incident THz pulse into the $\sigma_{\pm}$ circular basis. The reflection coefficient of each $\sigma_{\pm}$ component was obtained thanks to a polarization-resolved detection. A description of the experimental setup and of the polarization-resolved THz measurements is reported in the Supporting Information.
The magnetic field was applied by a large static magnet located behind the sample and was calibrated with a Hall probe. Homogeneity of the magnetic-field and the uncertainty on its absolute magnitude were better than $3\%$ across the THz spot size.

\subsection*{Numerical simulations}

Numerical simulations of the cavities presented in Fig. 2b were performed with the Rigorous Coupled Wave Analysis method \cite{Moharam_1995,Hugonin_2021}. The optical and geometrical parameters used in the simulations are determined from experimental measurements performed on the samples (see Supporting Information). The permittivity of Gold at THz frequencies was taken from \cite{Todorov_2010} and the permittivity of $Si_3N_4$ from \cite{Aupiais_2023}.

\section*{Acknowledgements}

Y.L. acknowledges support from Agence Nationale de la Recherche (grant n$^{\circ}$ ANR- 17-CE30-0011-11) and Labex PALM (grant n$^{\circ}$ ANR-10-LABX-0039-PALM).

\bibliography{Biblio_chiral_cavities}

\end{document}


\noindent 
\textbf{Contents:} 
15 pages, 6 Figures, 1 Table\\

\section{THz magneto-plasmonic properties of bulk InSb}
Under the application of a magnetic field $\boldsymbol{B}=B\boldsymbol{z}$ perpendicular to the interface of the plasmonic medium, the dielectric tensor of the medium acquires off-diagonal components and writes \cite{Palik_1970}:\\
\begin{equation}
	\overline{\overline{\epsilon \left(\nu \right)}}=\left( \begin{array}{ccc}
		{\epsilon }_{xx}\left(\nu \right) & {\epsilon }_{xy}\left(\nu \right) & 0 \\ 
		{-\epsilon }_{xy}\left(\nu \right) & {\epsilon }_{xx}\left(\nu \right) & 0 \\ 
		0 & 0 & {\epsilon }_{zz}\left(\nu \right) \end{array}
	\right)
\end{equation}
\linebreak
\noindent with ${\epsilon }_{xx}\left(\nu \right)={\epsilon }_{\infty }\left(1-\frac{{{\nu }_p}^2}{\nu }*\frac{(\nu +i\mathrm{\Gamma })}{{(\nu +i\mathrm{\Gamma })}^2-{{\nu }_c}^2}\right),{\epsilon }_{xy}\left(\nu \right)=-i\frac{{\epsilon }_{\infty }*{{\nu }_p}^2}{\nu }*\frac{{\nu }_c}{{(\nu +i\mathrm{\Gamma })}^2-{{\nu }_c}^2}$ and ${\epsilon }_{zz}\left(\nu \right)={\epsilon }_{\infty }\left(1-\frac{{{\nu }_p}^2}{\nu (\nu +i\mathrm{\Gamma })}\right)$. Here, $\nu$, $\nu_p$, $\nu_c$, $\Gamma$ and $\epsilon_\infty$ are the frequency, plasma frequency, cyclotron frequency, carrier scattering rate and background permittivity of the material, respectively. 

\begin{figure*}[ht]
	\centering 
	\includegraphics[width=0.7\textwidth]{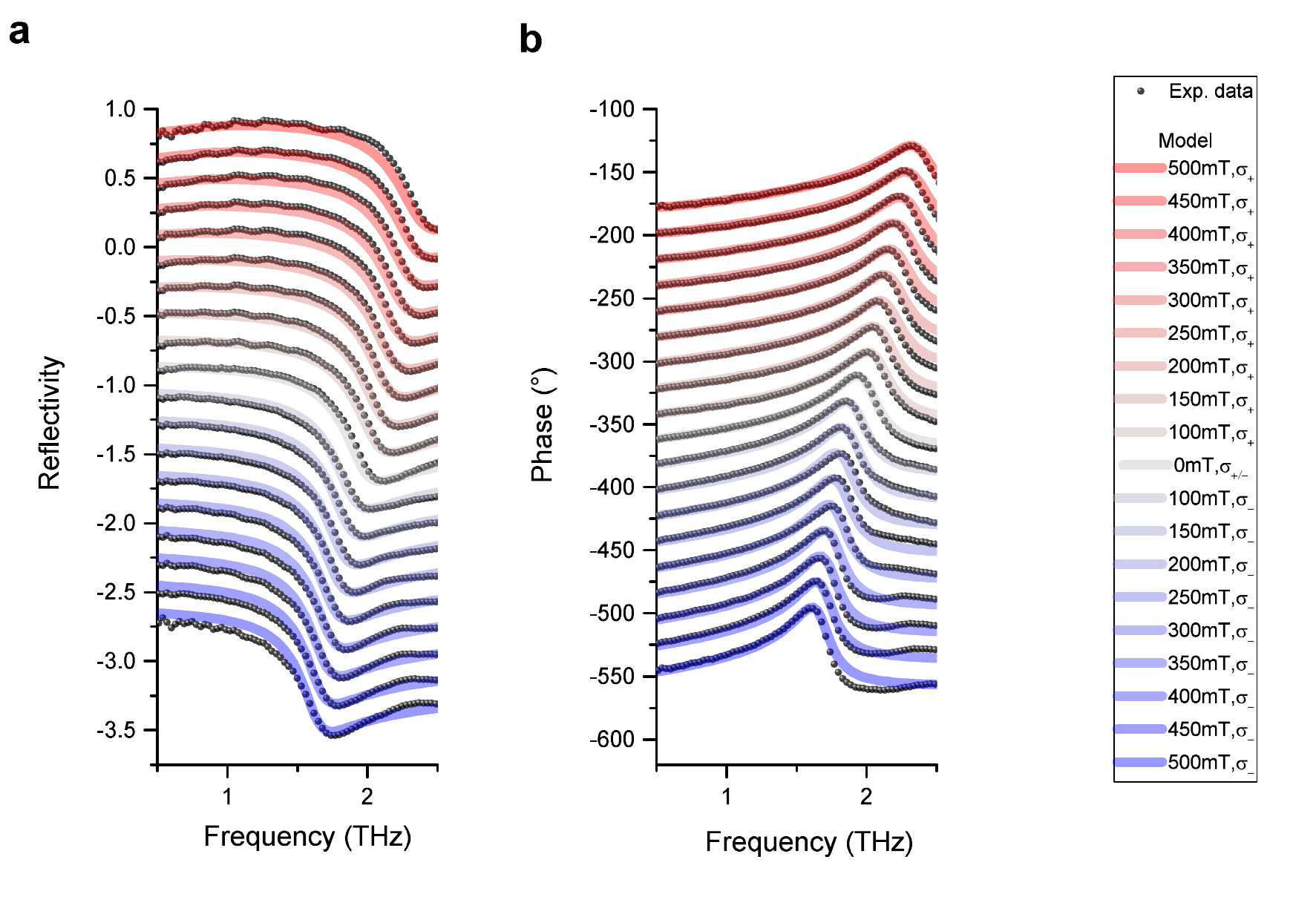}
	\caption{Polarization-resolved experimental THz reflectivity spectra  of bulk InSb as a function of magnetic field (black dots) together with the bests fits derived from the theoretical model for bulk parameters determination (thick coloured lines). \textbf{a} Power reflectivity. \textbf{b} Phase of the reflected THz field. Spectra are offset for clarity. }
	\label{fig:1}
\end{figure*}

\noindent The magnetic field, if spatially homogeneous, acts in the spin sector (i.e. the polarization) of the EM-field only. For a plane wave interacting with such a medium at normal incidence, the so-called Faraday configuration, normal modes correspond left- and right- circularly polarized (LCP/RCP) photons ${\widetilde{\boldsymbol{E}}}_{{\sigma }_{\pm }}\boldsymbol{=}e^{ik_zz}e^{-i\omega t}\left.\left|{\sigma }_{\pm }\right.\right\rangle $. Their interaction with the THz plasma is determined by the two different magneto-plasmonic permittivities in the circular basis: ${\epsilon }_{\sigma_{\pm} }\left(\nu \right)={\epsilon }_{xx}\left(\nu \right)\pm i{\epsilon }_{xy}\left(\nu \right)={\epsilon }_{\infty }\left(1-\frac{{{\nu }_p}^2}{\nu }*\frac{1}{\nu \pm {\nu }_c+i\mathrm{\Gamma }}\right)$.\\
Polarization-resolved THz spectra are acquired at room temperature as a function of magnetic-field $B$ for $\sigma_{\pm}$ polarizations. Fits of the experimental spectra according to this theoretical model are performed simultanesouly on the two quadratures of the reflection coefficient in order to determine the material parameters. A comparison between experimental data and the best fits is reported in Fig. S\ref{fig:1}.
From the $B=0mT$ spectrum, we determine $\epsilon_\infty=17.5$, $\nu_p=1.97THz$ and $\Gamma=0.27THz$. These parameters are kept independent of the magnetic-field and the spectra at $B\neq0mT$ are then fitted to determine the magnetic-field dependence of the cyclotron frequency $\nu_c$ that is reported in Fig. 1b of the main text.
From the linear relationship obtained between the cyclotron frequency and the magnetic-field ($\nu_c=eB/2\pi m^*$), we determine an electron effective mass of $m^*=0.019m_0$ ($m_0$ being the electron mass), consistent with previous reports \cite{Chochol_2017}.

\section{Perturbation theory of circular surface plasmonic cavities}
In order to understand the mode splitting of the cavity resonances arising from the application of a magnetic field, we apply a modal perturbation theory to the unperturbed normal modes of the surface plasmon resonator. For the sake of discussion, we consider a hermitian system where we neglect dissipation and where the cavity is isolated from its environnement. This amounts to a radiationless boundary condition where the normal component of the electric field is maximum on the edges of the circular patch (Neumann boundary condition) and is a very good approximation according to the simulations of the mode profiles shown in Fig. 2d of the main text. Under these assumptions, electric field of the cavity modes in the absence of an applied magnetic field can be written inside the plasmonic medium and in cylindrical coordinates as \cite{Minkowski_2014, Bigourdan_2016}:
\begin{equation} \label{GrindEQ__1_} 
{\widetilde{\boldsymbol{E}}}_{n,L}\boldsymbol{=}\left[iJ_{L+1}\left(k_{n,L}r\right)\left.\left|{\sigma }_+\right.\right\rangle -iJ_{L-1}\left(k_{n,L}r\right)\left.\left|{\sigma }_-\right.\right\rangle -k_{n,L}{\chi }_{n,L}J_L\left(k_{n,L}r\right)\left.\left|{\sigma }_z\right.\right\rangle \right]e^{iL\theta }e^{-\ \frac{z}{{\chi }_{n,L}}}e^{-i\omega t} 
\end{equation} 
Here,$\ J_{\alpha }(x)$ is the ${\alpha }^{th}$ Bessel function of first kind, $k_{n,L}$ is the plasmon wavector,$\ {\chi }_{n,L}$ is the penetration depth of the plasmon inside the plasmonic medium. Basis function are : $\left.\left|{\sigma }_+\right.\right\rangle =(1,i,0)$, $\left.\left|{\sigma }_-\right.\right\rangle =(1,-i,0)$ and $\left.\left|{\sigma }_z\right.\right\rangle =(0,0,1)$ written in the cylindrical basis $({\boldsymbol{e}}_{\boldsymbol{r}},{\boldsymbol{e}}_{\boldsymbol{\theta }},{\boldsymbol{e}}_{\boldsymbol{z}})$. The plasmon wavector $k_{n,L}\ $is determined by boundary conditions at the resonator edges to be such that: $k_{n,L}s/2$=${\mathrm{\Lambda }}_{n,L}$ where ${\mathrm{\Lambda }}_{n,L}$ is the n${}^{th}$ zero of the first derivative of the Bessel function $J_L(x)$. Normal modes are characterized by an orbital angular moment $L\ (L=\ \dots ,-2,-1,0,1,2,\dots )\ $ and a radial number $n\ (n=1,2,\ \dots )$.
\noindent Applying a modal perturbation theory to these modes allows to determine the relative frequency shifts $\delta {\nu }_{n,L}/{\nu }_{n,L}$ of the resonance frequencies ${\nu }_{n,L}$:
\[\frac{\delta {\nu }_{n,L}}{{\nu }_{n,L}}=\frac{\int_{V}{{\widetilde{\boldsymbol{E}}}^{\boldsymbol{*}}_{n,L}\overline{\overline{\mathrm{\Delta }\epsilon \left(\nu_{n,L} \right)}}{\widetilde{\boldsymbol{E}}}_{n,L}d\boldsymbol{r}}}{\int_{V}{\epsilon \left(\nu_{n,L} \right){\widetilde{\boldsymbol{E}}}^{\boldsymbol{*}}_{n,L}{\widetilde{\boldsymbol{E}}}_{n,L}d\boldsymbol{r}}}\] 
Here, $\overline{\overline{\mathrm{\Delta }\epsilon \left(\nu \right)}}$ corresponds to the variation of the dielectric tensor upon application of a magnetic field, it is non null only inside the plasmonic medium and writes to first order in $\nu_c$ and in cartesian coordinates:
\begin{equation*}
\overline{\overline{\mathrm{\Delta }\epsilon \left(\nu \right)}}=\left( \begin{array}{ccc} 0 & -i\frac{{\epsilon }_{\infty} \nu_p^2 \nu_c }{\nu^3} & 0 \\ 
i\frac{{\epsilon }_{\infty} \nu_p^2 \nu_c }{\nu^3} & 0 & 0 \\ 
0 & 0 & 0 \end{array}
\right)
\end{equation*}
and $W=\int_V{\epsilon \left({\nu }_{n,L}\right){\widetilde{\boldsymbol{E}}}^{\boldsymbol{*}}_{n,L}{\widetilde{\boldsymbol{E}}}_{n,L}}$ corresponds to the energy stored in the whole volume $V$ of the resonator, $\epsilon \left({\nu }\right)$ being the spatially independent and isotropic permittivity of the system in the absence of magnetic field.
This leads to:
\[\frac{\delta {\nu }_{n,L}}{{\nu }_{n,L}}={\nu }_c\frac{{{\nu }_p}^2{\epsilon }_{\infty}{\chi }_{n,L}}{2{{\nu }_{n,L}}^3}{\left(\int^{s/2}_0{r}\left({\left|J_{L+1}\left(k_{n,L}r\right)\right|}^2-{\left|J_{L-1}\left(k_{n,L}r\right)\right|}^2\right)dr\right)}/{W}\propto {\nu }_c\left(\int^{s/2}_0{r{\mathrm{\ }\mathrm{\Delta }\sigma }_{n,L}\left(r\right)dr}\right)\] 
where $\mathrm{\Delta }{\sigma }_{n,L}\left(r\right)={\left|J_{L+1}\left({k }_{n,L}r\right)\right|}^2-{\left|J_{L-1}\left({k}_{n,L}r\right)\right|}^2$. \\ \\ Written in this form, it is clear that $\mathrm{\Delta }{\sigma }_{n,L}\left(r\right)$ corresponds to the difference in the spin density between $\sigma_+$ (density given by ${\left|J_{L+1}\left({k}_{n,L}r\right)\right|}^2$) and $\sigma_-$ (density given by ${\left|J_{L-1}\left({k}_{n,L}r\right)\right|}^2$) photons.  $\mathrm{\Delta }{\sigma }_{n,L}\left(r\right)$ is plotted in Fig. 1d of the main text.
\\The presence of the form factor $\int^{s/2}_0{r{\mathrm{\ }\mathrm{\Delta }\sigma }_{n,L}\left(r\right)dr}$ hence indicates that the amplitude of the mode splitting is controlled by the integrated relative spin densities of the $\sigma_{\pm}$ polarized photons inside the plasmonic medium where the perturbation is induced. This term is maximized for $(n,L)=(1,\pm1)$ and decreases for increasing $n$ and $L$'s, so that the mode splitting is largest for the $(n,L)=(1,\pm1)$ cavity modes.

\section{Geometrical parameters of the samples}

The geometrical parameters $s$ and $a$ of the samples are determined from optical microscopy measurements. The insulator thickness $d$ is determined from ellipsometry measurements. These parameters are reported in Tab. S1.

\begin{table*}[h!]
	\center
	\begin{tabular}{| c | c | c | c | c | c | c |}
		\hline
		& Sample 1 & Sample 2 & Sample 3  \\ \hline
		$s$& $42.5 \ \mu m $ & $32.0 \ \mu m $ & $21.3 \ \mu m $\\ \hline
		$a$ & $14.5 \ \mu m $ & $11.3 \ \mu m $ & $7.9 \ \mu m $\\ \hline
		$d$ & $2.2 \ \mu m$ & $2.2 \ \mu m$ & $2.2 \ \mu m$ \\ \hline
	\end{tabular}
	\caption{Geometrical parameters of the samples}
	\label{tab:1}
\end{table*}
The values of these parameters have been rounded in the main text for simplicity.

\section{Experimental setup}
\begin{figure*}[h!]
	\centering 
	\includegraphics[width=1\textwidth]{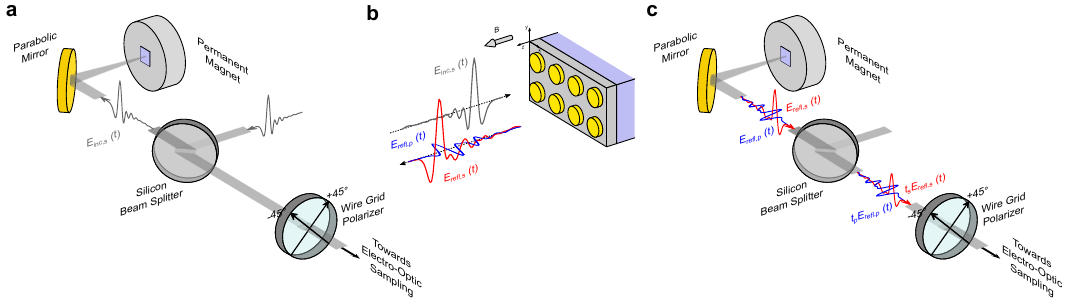}
	\caption{Experimental setup for the $\sigma_{\pm}$ polarization-resolved THz measurement \textbf{a} A vertically polarized THz pulse $E_{inc,s}(t)$ is impinging onto the sample at normal incidence  \textbf{b} The THz pulse is reflected from the sample with components $E_{refl,s}(t)$ and $E_{refl,p}(t)$ in the vertical and horizontal polarizations respectively. \textbf{c} The two orthogonal components of the THz pulse are transmitted through a Si beam splitter and emerge from it with amplitudes $t_s E_{refl,s}(t)$ and $t_p E_{refl,p}(t)$  ($t_s/t_p$ being the transmission coefficients of the Si beam splitter for s- and p- polarizations, respectively) before they are read out by a cross-polarizer oriented at $\pm 45\degree$ and an electro-optic detection.}
	\label{fig:2}
\end{figure*}

\begin{figure*}[h!]
	\centering 
	\includegraphics[width=1\textwidth]{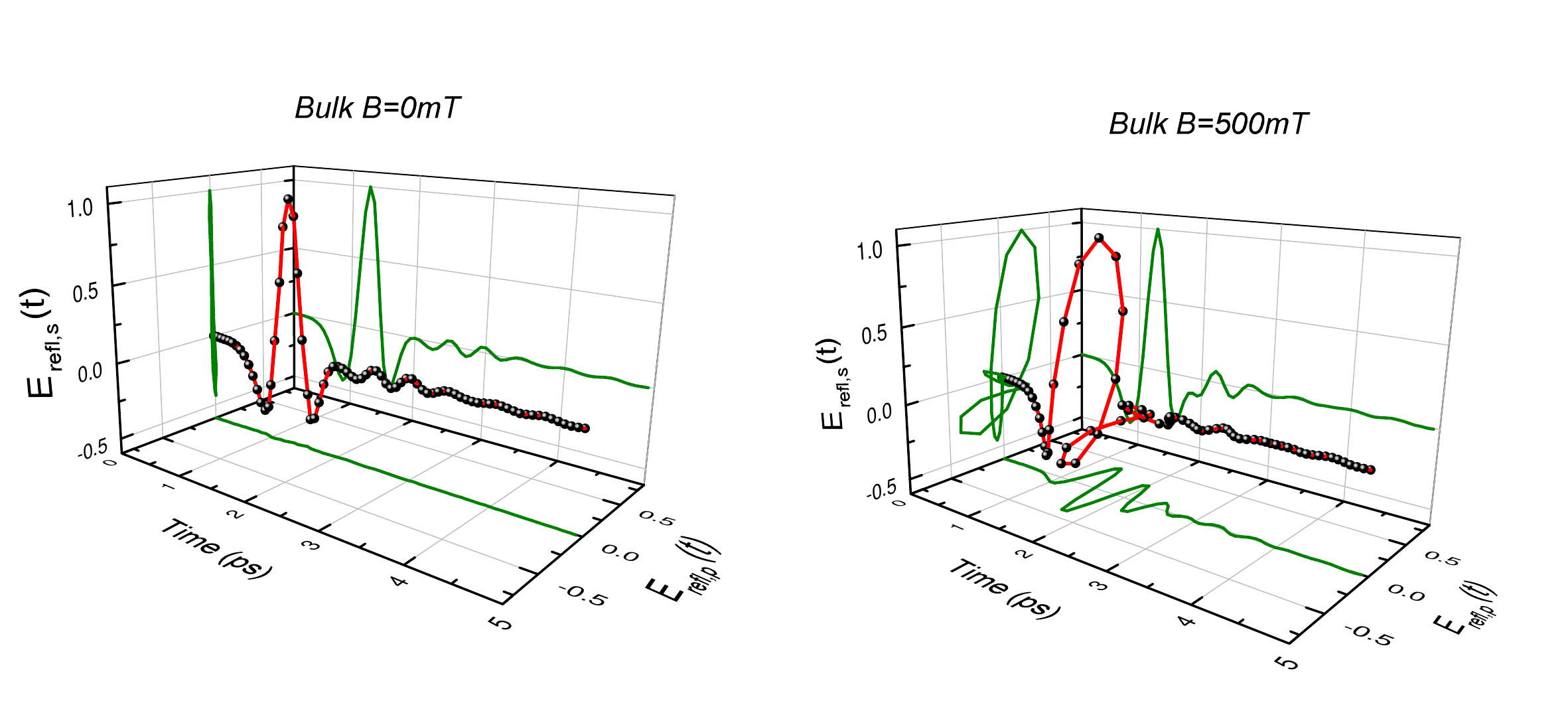}
	\caption{Retrieval of the s- and p-components of the THz pulse reflected from the bulk material after measurement from the cross-polarizer and the electro-optic detection and correction from the transmittion coefficients $t_s/t_p$ of the Si beam splitter (see text for details). Left: THz pulse reflected from bulk InSb in the absence of a magnetic field; Right: THz pulse reflected from bulk InSb subjected to a magnetic field $B=500mT$.}
	\label{fig:3}
\end{figure*}

The experimental setup is depicted in Fig. S\ref{fig:2}. \\ 
\noindent Normal incidence reflectivity measurement are performed using a Si beam splitter placed at 45\degree \space with respect to the sample's normal.
Incident THz pulses are linearly polarized in the vertical polarization $E_{inc,s}(t)$ ($s$-polarization, Fig. S\ref{fig:2}a). They hence consist in a coherent superposition of $\sigma_{\pm}$ polarized pulses $E_{inc,s} (t)=\frac{1}{\sqrt{2}}(E_{inc,\sigma_{+}}(t) + E_{inc,\sigma_{-}}(t))$ with $E_{inc,\sigma_{+}}(t)=E_{inc,\sigma_{-}}(t)=E_{inc,s} (t)/\sqrt{2}$. The THz pulses reflected off the sample ($E_{refl,s}(t)$ and $E_{refl,p}(t)$ for vertical and horizontal polarizations respectively, Fig. S\ref{fig:2}b) are first transmitted through the Si beam splitter with respective amplitudes $t_s E_{refl,s}(t)$ and $t_p E_{refl,p}(t)$ ($t_{s/p}$ being transmission amplitudes of the $s/p$ polarizations of the Si beam splitter) before they reach a crossed polarizer and electro-optic detection (Fig. S\ref{fig:2}c). The crossed polarizer detection, achieved with wire-grid polarizers rotated at either $+45 \degree$ or $-45 \degree$ from the vertical direction, is used to measure the two orthogonal polarizations of the reflected THz pulses according to $ E_{refl,s}(t)= \frac{1}{\sqrt{2} t_{s}} (E_{+45 \degree}(t) + E_{-45 \degree}(t))$ and $E_{refl,p}(t)= \frac{1}{\sqrt{2} t_{p}} (E_{+45 \degree}(t) - E_{-45 \degree}(t))$. Finally, the reflected THz pulses are reconstructed in the circular basis: $E_{refl,\sigma_{\pm}}(t)=\frac{1}{\sqrt{2}} (E_{refl,s}(t) \pm i E_{refl,p}(t))$. \\ 
\noindent The polarization resolved reflection coefficients $r_{\sigma_{\pm}}(\nu)$ are determined from Fourier transforms of the incident and reflected pulses: $r_{\sigma_{\pm}}(\nu)=E_{refl,\sigma_{\pm}}(\nu)/E_{inc,\sigma_{\pm}}(\nu)$. Here, $E_{inc,\sigma_{\pm}}(\nu)$ is obtained in a separate mesurement from the reflection off a plane gold surface located at the sample's position.
An example of such orthogonal basis decomposition obtained on the bulk material is shown in Fig. S\ref{fig:3}. 

\section{Experimental determination of cavity parameters}
\subsection{Model of reflectivity}
In the spirit of the non-hermitian scattering matrix and quasi-normal modes formulations of Maxwell's equations \cite{Alpeggiani_2017,Lalanne_2018,Zhang2020,Benzaouia_2021,Sauvan_2022}, the complex-valued reflection coefficient (denoted $\tilde{r}_{\sigma_{\pm}}(\nu,B)$) of our samples can be understood as originating from the background reflectivity of a non structured $vacuum/Si_3N_4/InSb$ planar multilayer (denoted $\tilde{r}_{\sigma_{\pm},0}(\nu,B)$) onto which the insertion of metallic patches leads to the emergence of 3 major resonances that we observe in the spectra: the cavity resonance, the intercavity resonance as well as the first order grating diffraction of waves into the bulk of InSb. The latter resonance originates from the periodic nature of the structure and arises for frequencies located in the transparency band of the plasma above the plasma frequency.
Hence, we physically model the reflectivity coefficient as follows:
\begin{equation}
	\tilde{r}_{\sigma_{\pm}}(\tilde{\nu},B)=\tilde{r}_{\sigma_{\pm},0}(\tilde{\nu},B)+\frac{\tilde{C_0}}{\tilde{\nu}-\tilde{\nu_0}}+\frac{\tilde{C_1}}{\tilde{\nu}-\tilde{\nu_1}}+\frac{\tilde{C_2}}{\tilde{\nu}-\tilde{\nu_2}}+\tilde{C_3}+\tilde{C_4} \tilde{\nu}+\tilde{C_5} \tilde{\nu}^2
	\label{model}
\end{equation}
Here, $(\tilde{\nu_0},\tilde{\nu_1},\tilde{\nu_2})$ and $(\tilde{C_0},\tilde{C_1},\tilde{C_2})$ correspond to the frequencies of the poles and coupling constants of, respectively, the cavity resonance, the intercavity resonance and the grating diffraction, whereas $\tilde{C_3}+\tilde{C_4} \tilde{\nu}+\tilde{C_5} \tilde{\nu}^2$ describe residual terms. The latter terms take into account the fact that the resonance associated with the grating diffraction is not a simple pole but a cut in the complex plane due to the continuum of propagating modes, as observed in our simulations (not shown). 
In this formula, all the parameters $\left\{\tilde{\nu_i},\tilde{C_i}\right\}$ are complex numbers and depend implicitely on the magnetic-field $B$ and the polarization $\sigma_{+}/\sigma_{-}$ of the THz light. Furthermore, the frequency $\tilde{\nu}$ is made complex-valued by analytical continuation of reflectivity into the complex-plane \cite{Popov_1986,Nevi_re_1995,Binkowski_2024}, allowing for a description of the resonance based on a pole-zero representation of the cavity parameters (see below). The poles of the grating diffraction resonance are kept fixed and independent of the magnetic field and polarization as observed experimentally ($\tilde{\nu_2}=2.27-0.2i,2.6-0.2i$ and $3.6-0.2i$ for the $42\mu m, 32 \mu m$ and $21\mu m$ cavity samples respectively).\\
The fits are performed simultaneously on the two quadratures of the reflectivity coefficient, such as amplitude and phase or its real and imaginary parts.
By performing RCWA simulations of our structures, we benchmarked this model by comparing together the poles and zeros determined for the cavity resonance in the following two cases: 1) by fitting the reflectivity spectra computed for real frequencies (i.e. the one accessible in an experimental context) and 2) by obtaining the exact poles and zeros from a search of the cavity resonances in the complex plane. The excellent agreement that we obtained between the two approaches, where real and imaginary parts of the poles and zeros determined from the fit are within 2-3\% of the values obtained from the search in the complex plane, validates this model. 
\subsection{Fits of the experimental data}
\begin{figure*}[h!]
	\centering 
	\includegraphics[width=1\textwidth]{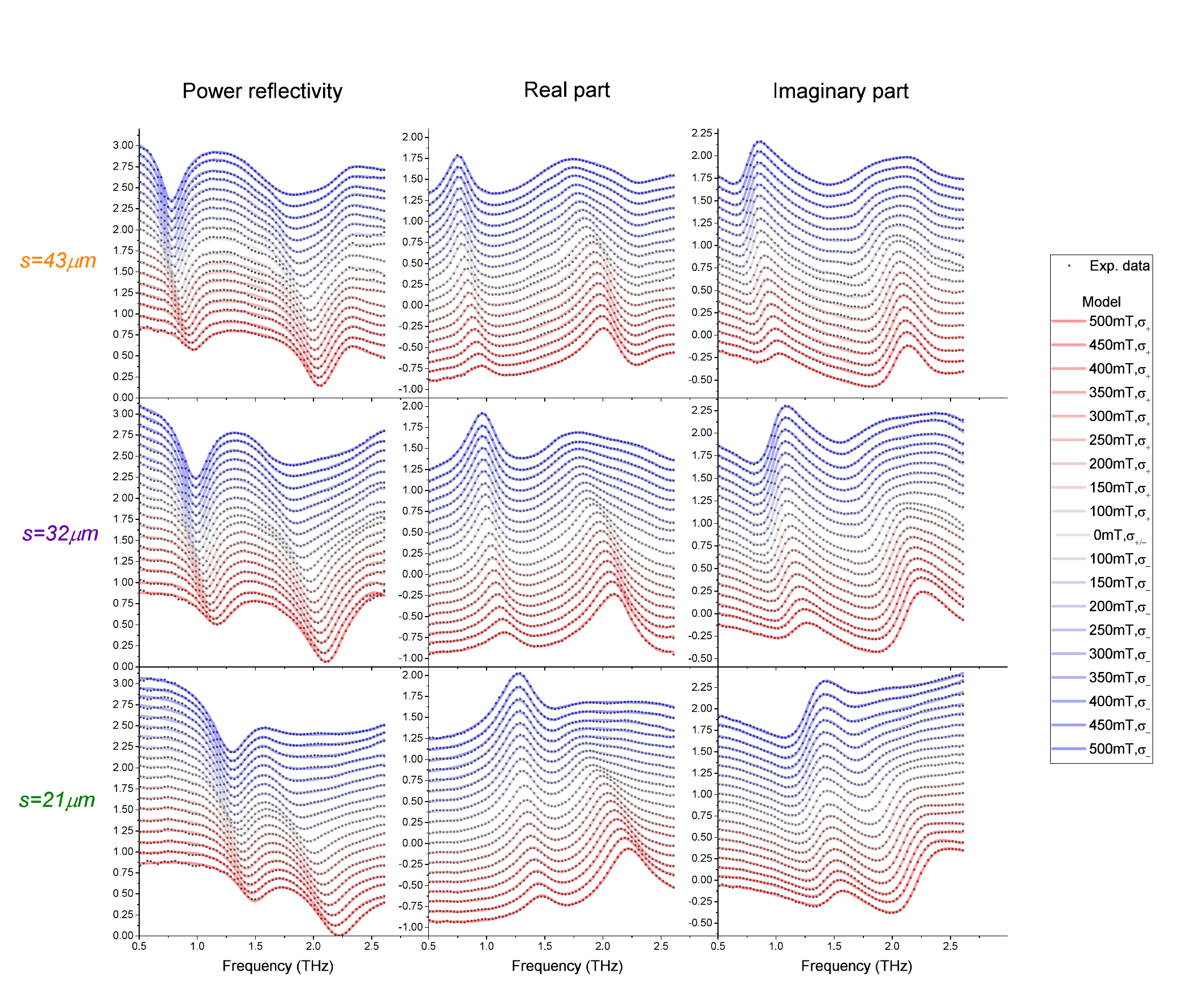}
	\caption{Polarization-resolved experimental THz reflectivity spectra of the three samples as a function of magnetic field (black dots) together with the best fits (coloured lines) obtained from the model of reflectivity eq. (\ref{model}). From top to bottom: $s=43\mu m, 32\mu m$ and $21\mu m$ cavity samples. From left to right: Power reflectivity $\left| \tilde{r}_{\sigma_{\pm}}(\nu,B) \right|^2$, real and imaginary parts of the reflection coefficient $\tilde{r}_{\sigma_{\pm}}(\nu,B)$. Spectra are offset for clarity.}
	\label{fig:4}
\end{figure*}

Equipped with this model, we use it to fit the experimental reflectivity spectra for the cavity parameters determination. In Fig. S\ref{fig:4}, we report the comparison between experimental data and the best fits obtained. We find that the model reproduces very well the experimental reflectivity spectra for all three samples, at all fields B and in both  $\sigma_{+}/\sigma_{-}$ polarizations.
\subsection{Pole-zero representation of the cavity parameters}
Formula (3) allows to determine the pole $\tilde{\nu_0}$ of the cavity resonance. While we determine the zero $\tilde{\nu}_{z0}$ of the resonance from the condition $\tilde{r}_{\sigma_{\pm}}(\tilde{\nu}_{z0},B)=0$ once all parameters are known, the zero does not explicitly appear in the formula.
To make the zero explicitly appear in this formula, we note that the reflectivity around the pole $\tilde{\nu}_0$ takes the form
$\tilde{r}_{\sigma_{\pm}}(\tilde{\nu},B)=\tilde{a}(\tilde{\nu})+\frac{\tilde{C_0}}{\tilde{\nu}-\tilde{\nu_0}}$ where $\tilde{a}(\tilde{\nu})$ is a background term that is a slowly varying function of $\tilde{\nu}$ around $\tilde{\nu}_0$. Using complex analysis, this can be re-written as:
\begin{equation} \tilde{r}_{\sigma_{\pm}}(\nu,B)=\tilde{a}(\tilde{\nu})*\frac{\tilde{\nu}-\tilde{\nu}_{z0}}{\tilde{\nu}-\tilde{\nu}_0}
\label{zero}
\end{equation}
where it is now made explicit that the reflectivity vanishes at the complex frequency $\tilde{\nu}=\tilde{\nu}_{z0}$ which defines the location of the zero of the resonance in the complex plane. Alternatively, it can also written as follows:
$\tilde{r}_{\sigma_{\pm}}(\tilde{\nu},B)=\tilde{a}(\tilde{\nu})*(1+\frac{\tilde{\nu}_0-\tilde{\nu}_{z0}}{\tilde{\nu}-\tilde{\nu}_{z0}})$.
This formula tells that when the pole and zero are displaced from each other ($\tilde{\nu}_0\neq\tilde{\nu}_{z0}$), the cavity resonance appears as a resonance modulating a background reflectivity given by $\tilde{a}(\tilde{\nu})$. We have verified that use of either of these formulas instead of formula (\ref{model}) leads to the same results for the determination of the poles and zeros of the cavity resonance, hence validating the robustness of the cavity parameters' determination. \\
Formula (\ref{zero}) allows to make the link with other more common formulations by physically expressing the poles and zero in terms of commonly used quantities such as radiative ($\Gamma_{r}$) and non radiative ($\Gamma_{\ohm}$) decay rates.
To see this, we recall the functional form of the reflection coefficient used for extracting resonance paramaters in metal/insulator/metal THz microcavities, as derived from temporal coupled mode theory and commonly written for real frequencies $\nu$ as \cite{Manceau_2013, Qu_2015}:
\begin{equation}
	\tilde{r}(\nu)=-1+\frac{i2\Gamma_r}{\nu-\nu_0+i(\Gamma_{\ohm}+\Gamma_r)}=-1*\frac{\nu-\nu_0+i(\Gamma_{\ohm}-\Gamma_r)}{\nu-\nu_0+i(\Gamma_{\ohm}+\Gamma_r)}
	\label{mim}
\end{equation}
This expression is equivalent to eq.(\ref{zero}) when we take $\tilde{a}(\nu)=-1$, $\tilde{\nu_{0}}=\nu_{0}-i(\Gamma_{\ohm}+\Gamma_r)$ and  $\tilde{\nu}_{z0}=\nu_{0}-i(\Gamma_{\ohm}-\Gamma_r)$.
Indeed, for metal/insulator/metal THz microcavities, the background reflectivity $\tilde{a}(\nu)$ is usely taken to be dispersionless and equal to -1 as it originates from the reflection upon the bottom metal where the THz electric field undergoes $\approx$100\% power reflection with a $\pi$ phase shift. We hence conclude that formula (\ref{zero}) can be considered as a generalization of eq.(\ref{mim}) that allows to treat our highly dispersive and non-hermitian plasmonic system. In particular, we note the following points:
\begin{itemize}
\item as opposed to eq.(\ref{mim}), eq. (\ref{zero}) allows to describe resonances whose real parts of the pole and zero may not be necessarily equal: $Re(\tilde{\nu}_0)\neq Re(\tilde{\nu}_{z0})$. This allows to describe assymetric, i.e. Fano-type, lineshapes of the resonances. In our analysis, we find that $\nu_{0}=Re(\tilde{\nu}_{0})\approx Re(\tilde{\nu}_{z0})$ in the whole range of parameters investigated and hence do not discuss this aspect further.
\item By identification of the two formulas, we ascribe the real and imaginary parts of the pole $\tilde{\nu_0}$ to, respectively, the resonance frequency  and (minus) the total decay rates: $Re(\tilde{\nu}_0)=\nu_{0}$, $Im(\tilde{\nu}_0)=-(\Gamma_{\ohm}+\Gamma_r)$. Similarly, the imaginary part of the zero quantifies the relative balance of the two decay rates ($Im(\tilde{\nu}_{z0})=-(\Gamma_{\ohm}-\Gamma_r)$) and characterizes the coupling condition of the resonator (Fig. 4a of the main manuscript): 
\begin{itemize}
	\item when $\Gamma_{\ohm}>\Gamma_r$ ($Im(\tilde{\nu}_{z0})<0$), the resonator is undercoupled
	\item when $\Gamma_{\ohm}=\Gamma_r$ ($Im(\tilde{\nu}_{z0})=0$), the resonator is critically coupled
	\item when $\Gamma_{\ohm}<\Gamma_r$ ($Im(\tilde{\nu}_{z0})>0$), the resonator is overcoupled
\end{itemize}
\end{itemize}
Let's also note that thanks to the phase information provided by the time-domain THz spectrocopy technique, we perform fits on the two quadratures of the reflection coefficient simultaneously (both the real and imaginary part). As opposed to fits performed solely on the power spectrum and which would give access to $\left| \Gamma_{\ohm}-\Gamma_r \right|$, this approach allows to determine ($\Gamma_{\ohm}-\Gamma_r)$ without any ambiguity regarding its sign. This is particularly relevant when determining accurately if the resonator is under- or over- coupled.
Finally, $\Gamma_{\ohm}$ and $\Gamma_{r}$ are determined from the knowledge of the pole and zero via: $\Gamma_{\ohm}=-Im(\tilde{\nu}_0+\tilde{\nu}_{z0})/2$ and $\Gamma_{r}=-Im(\tilde{\nu}_0-\tilde{\nu}_{z0})/2$.\\
Based on this analysis, we report in Fig. S5 all the cavity parameters and their magnetic field dependence: real parts of the poles and zeros as well as the ohmic and radiative decay rates.
\begin{figure*}[h!]
	\centering 
	\includegraphics[width=1\textwidth]{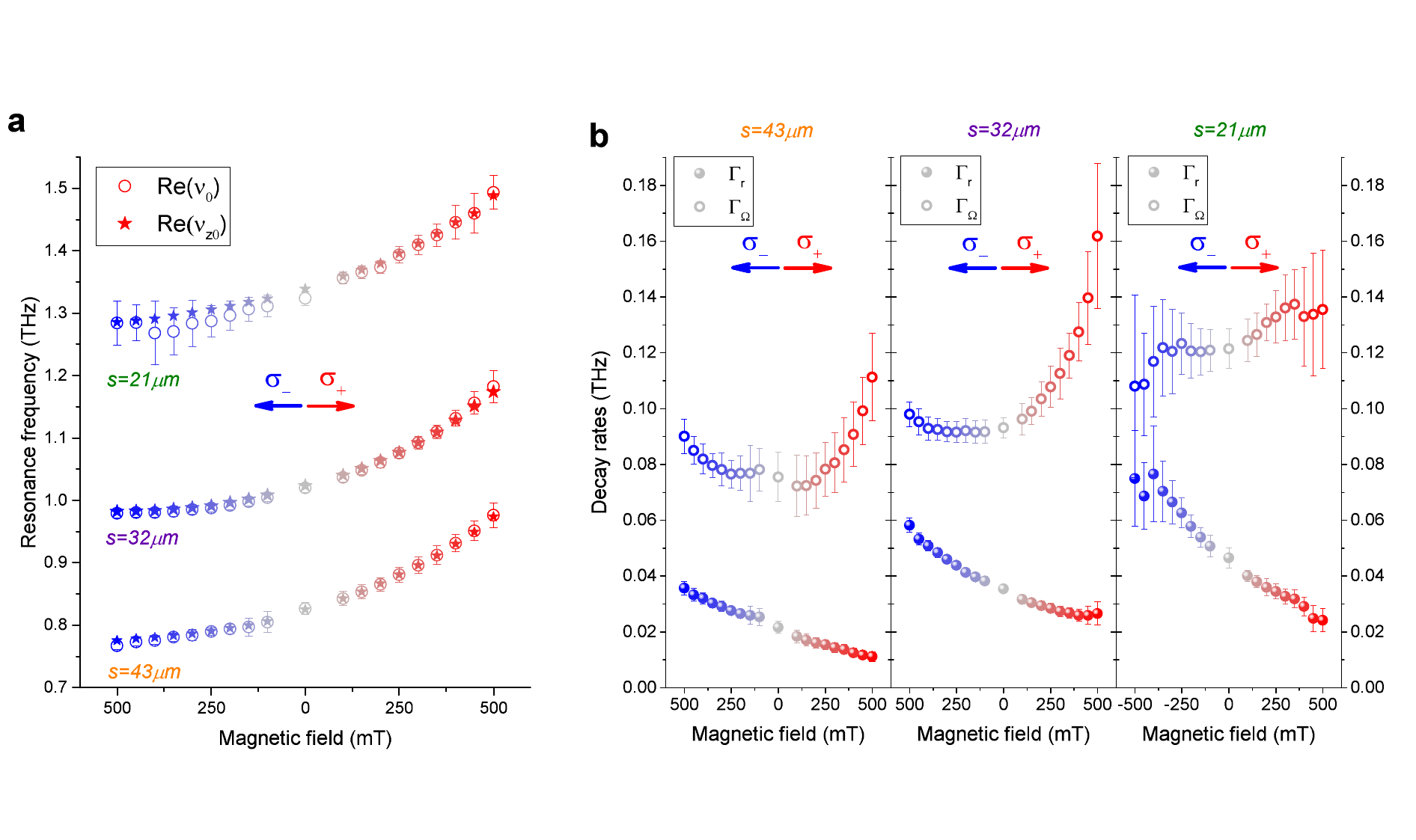}
	\caption{Cavity parameters obtained from fits of the reflectivity and their evolution as a function of magnetic field for $\sigma_{\pm }$ polarizations. \textbf{a} Cavity resonance frequency: real parts of the pole (open circles) and zero (stars) of the cavity resonance. Error bars represent 95\% confidence intervals on the real part of the pole. \textbf{b} Non-radiative ($\Gamma_{\ohm}$, open circles) and radiative ($\Gamma_{r}$, filled circles) decay rates. Error bars represent 95\% confidence intervals. Note that the graphs are double sided with respect to the magnetic field. Left side:  $\sigma_{-}$ polarization; Right side:  $\sigma_{+}$ polarization. }
	\label{fig:5}
\end{figure*}
\clearpage
\section{Optical activity of chiral THz metasurfaces}
In Fig. S6, we report the optical activity of the three metasurfaces together with that of the bulk plasmonic material for comparison. We note in particular the wide tunability of the optical activity that can be engineered with the metasurfaces: 1) spectrally within the whole opacity band of the plasma, even well below the plasma frequency, via the geometrical tuning of the cavity resonance frequency and 2) in its amplitude by adjusting the strenght of the magnetic field.
\begin{figure*}[h!]
	\centering 
	\includegraphics[width=1\textwidth]{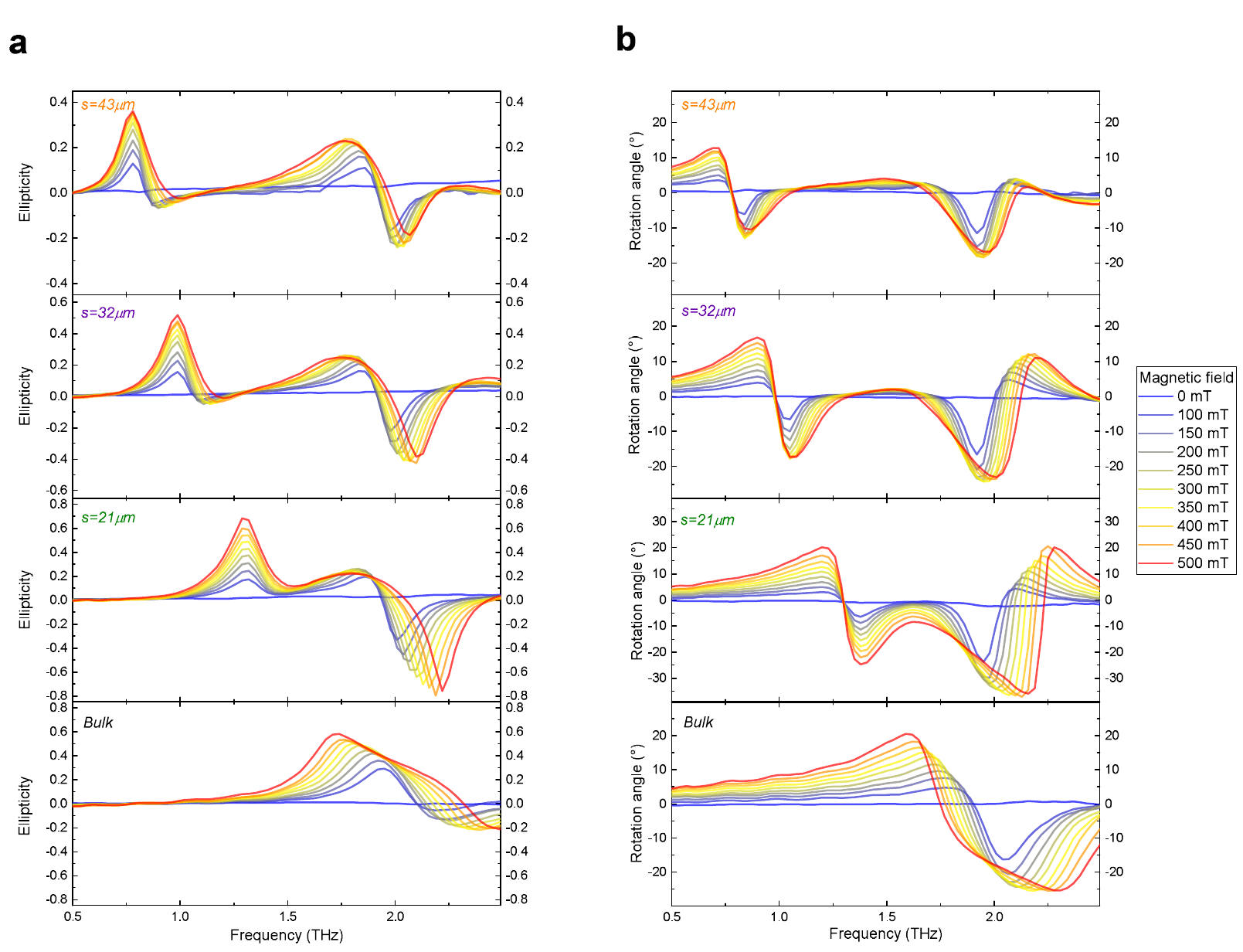}
	\caption{Comparison of the optical activity between the metasurfaces samples and the bulk plasmonic material. \textbf{a} Ellipticity \textbf{b} Optical rotation. See main text for the definition of these two quantities.}
	\label{fig:6}
\end{figure*}
\clearpage

\bibliography{Biblio_chiral_cavities}